\documentclass[11pt]{article}

\usepackage{amsmath}
\usepackage{graphicx}
\usepackage{amsfonts}
\usepackage{amssymb}
\usepackage{epsfig}
\usepackage{color}
\usepackage{psfrag}

\newcommand{\EQ}{\begin{equation}}
\newcommand{\EN}{\end{equation}}

\begin{document}
\setcounter{page}{0} \topmargin 0pt
\renewcommand{\thefootnote}{\arabic{footnote}}
\newpage
\setcounter{page}{0}

\begin{titlepage}

\begin{flushright}
NSF-KITP-06-62
\end{flushright}

\vspace{0.5cm}
\begin{center}
{\Large {\bf Holomorphic Parafermions in the Potts model and SLE}}\\

\vspace{2cm}
{\large V. Riva$^{a,b}$ and J. Cardy$^{a,c}$} \\
\vspace{0.5cm} {\em $^{a}$Rudolf Peierls Centre for Theoretical
Physics, 1 Keble Road, Oxford OX1 3NP, UK} \\\vspace{0.3cm} {\em
$^{b}$Wolfson College, Oxford} \\\vspace{0.3cm} {\em $^{c}$All
Souls College, Oxford}

\end{center}

\vspace{1cm}

\begin{abstract}

\noindent We analyse parafermionic operators in the $Q$--state
Potts model from three different perspectives. First, we
explicitly construct lattice holomorphic observables in the
Fortuin-Kasteleyn representation, and point out some special
simplifying features of the particular case $Q=2$ (Ising model).
In particular, away from criticality, we find a lattice
generalisation of the massive Majorana fermion equation. We also
compare the parafermionic scaling dimensions with known results
from CFT and Coulomb gas methods in the continuum. Finally, we
show that expectation values of these parafermions correspond to
local observables of the SLE process which is conjectured to
describe the scaling limit of the $Q$--state Potts model.

\vspace{2cm}

\hrulefill

E-mail addresses: j.cardy1, v.riva1@physics.ox.ac.uk

\end{abstract}

\end{titlepage}

\newpage

\section{Introduction}
\setcounter{equation}{0}

Among the special features of two--dimensional physics, one of the
most intriguing is semi--locality, which is marked by the
appearance of a complex phase in the correlation function of two
or more excitations once their positions have been exchanged. The
study of this type of excitations, called parafermions and
characterized by fractional spin, has found fruitful applications
in diverse areas of theoretical physics. Parafermions are rather
well understood in critical systems in the continuum limit, since
the discrete symmetries they are associated to combine with
conformal invariance and permit an extensive classification of the
so--called parafermionic Conformal Field Theories. The first of
such theories, constructed by Fateev and Zamolodchikov
\cite{zamfat}, describes the critical point of certain lattice
models with $\mathbb{Z}_N$ symmetry, which reduce to the Ising and
$3$--state Potts models for $N=2$ and $N=3$, respectively. The
parafermionic fields represent the chiral conserved currents
associated to the extra symmetry, the conservation laws being
expressed in complex coordinates as a holomorphicity condition on
the parafermions. In the Ising case, the parafermionic current is
precisely the chiral component of the Majorana fermion that is
known to describe the model at criticality and in the neighbouring
scaling regime, and the holomorphicity condition is deformed, away
from criticality, by terms proportional to the fermion mass.

In this paper, we analyse parafermionic observables in the
$Q$--state Potts model, both on the lattice and in the continuum
limit. With respect to the literature on the subject, our approach
provides a new explicit proof of holomorphicity on the lattice,
and the first analysis of its relation to Stochastic Loewner
Evolution (SLE) in the continuum limit.

On the lattice, we adopt the Fortuin--Kasteleyen (FK)
representation and define the parafermions in terms of the
corresponding gas of loops. This allows us to show that, for every
value of $Q$, there is a particular value of the spin at which the
parafermionic correlation functions are purely holomorphic (in a
lattice sense) at criticality. The resulting scaling dimensions
agree with CFT and Coulomb gas results in the continuum.

It is conjectured that the continuum limit of the $Q$--state Potts
model in the FK representation is described by SLE$_{\kappa}$ with
$\sqrt{Q}\,=\,-2\cos\left(4\pi/\kappa\right)$, $\;4<\kappa<8$. We
provide further evidence for this correspondence by constructing
SLE observables which are holomorphic, have the appropriate
scaling dimension and satisfy the same boundary conditions as the
lattice parafermions mentioned above. Finally, we explicitly
discuss the relation between parafermions and SLE on the lattice
in a particular example.

Our work, although carried out independently, has some overlap
with that of the programme of Smirnov\cite{SmirnovICM} for proving
that the scaling limit of certain curves in lattice O$(n)$ and
$Q$-state Potts models is SLE with the appropriate value of
$\kappa$ (which has been pushed to completion by Smirnov for the
Ising case ($n=1$, $Q=2$)). Our emphasis, however, is on the
lattice holomorphicity of the multi-point correlations of the
parafermions, and their relations to earlier physics CFT
literature, rather than on detailed considerations of the
one-point function in an arbitrary domain which are required to
prove convergence to SLE. That fact that analogous holomorphic
observables exist in SLE was already pointed out by B.~Doyon and
the present authors in an earlier paper\cite{Tbulk}.

The paper is organized as follows. In Section\,\ref{sectionPotts}
we briefly review the FK representation of the Potts model. In
Section\,\ref{sectlattice} we define the parafermionic operators
of spin $p$, and we show that their correlation functions are
purely holomorphic at a particular value of $p$. Special features
of the Ising model are discussed in Section\,\ref{sectising}.
Section\,\ref{sectcontlim} presents the continuum results, first
the ones already known from Coulomb gas and CFT methods, then the
new prediction obtained by SLE. Finally, in
Section\,\ref{sectrelSLElatt} we discuss the relation between the
SLE and lattice results, and we conclude the paper in
Section\,\ref{sectconclusions}.

\section{Q-state Potts model}\label{sectionPotts}
\setcounter{equation}{0}

\begin{figure}[h]
\begin{center}
\psfig{figure=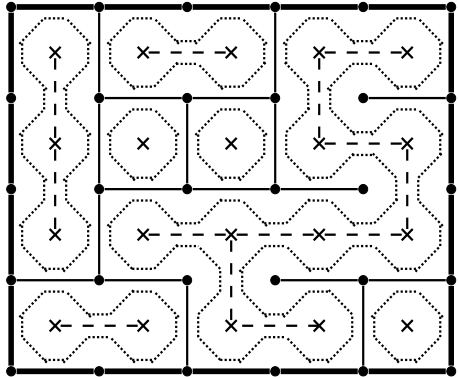,height=5cm,width=6cm}
\end{center}
\caption{A configuration in the FK representation of the Potts
model with wired boundary conditions (thick lines). Dots and
crosses represent sites of the original and dual lattice,
respectively.}\label{FKconfigwired}
\end{figure}

Let us consider the $Q$--state Potts model on the square lattice,
at each site of which is associated a spin variable
$S_i\in\{0,1,...,Q-1\}$. The partition function is
\begin{equation}\label{originalZ}
Z\,=\,\sum\limits_{\{S\}}\,e^{\sum\limits_{<i,j>}\,J\,\delta(S_i,S_j)}
\,=\,\sum\limits_{\{S\}}\,\prod\limits_{<i,j>}\,[1+u\,\delta(S_i,S_j)]\;,\qquad\qquad
u=e^{J}-1\;.
\end{equation}
By expanding the product in (\ref{originalZ}), one obtains the
Fortuin--Kasteleyn (FK) representation \cite{FK}
\begin{equation}\label{FKZ}
Z\,=\,\sum\limits_G\,u^b\,Q^c
\end{equation}
where $G$ is any subgraph of the original domain, consisting of
all the sites and some bonds placed arbitrarily on the lattice
edges, $b$ is the number of bonds in $G$, and $c$ is the number of
clusters of connected sites into which the bonds partition the
lattice (FK clusters, which also include single sites).
Fig.\,\ref{FKconfigwired} shows a configuration in this expansion,
where wired boundary conditions have been assigned (i.e.
$u=\infty$ on the boundary edges). Alternatively, one could assign
$u=0$ to the boundary edges, which corresponds to free boundary
conditions on the Potts spins.

\vspace{0.5cm}

Although the original formulation of the model requires $Q$ to be
a positive integer, the FK representation (\ref{FKZ}) allows one
to interpret $Q$ as taking arbitrary complex values. The model is
known to have a critical point with a diverging correlation length
at the self-dual value of $u$ for $Q$ real and $0<Q\leq 4$.

The partition function (\ref{FKZ}) can be also expressed as that
of a gas of fully packed loops on the medial lattice (the dashed
lines in Fig.\,\ref{FKconfigwired}), whose number is $d=l+c$ for
each given $G$, where $l$ is the number of independent loops in
$G$.

By using Euler's relation
\begin{equation}\label{euler}
c-l\,=\,N-b\;,
\end{equation}
where $N$ is the number of vertices in $G$ (i.e. the total number
of vertices in the lattice), the partition function can be
rewritten as
\begin{equation}\label{FPLZ}
Z\,=\,Q^{\frac{N}{2}}\sum\limits_G\,\left(\frac{u}{\sqrt{Q}}\right)^b\,\sqrt{Q}^{\,d}\;.
\end{equation}
By reexpressing (\ref{FPLZ}) in terms of the dual lattice and
imposing self duality at the critical point, one extracts the
critical value
\begin{equation}\label{uc}
u_c\,=\,\sqrt{Q}\;,
\end{equation}
which implies that the weights for the FK clusters at criticality
are equivalent to counting each loop on the medial lattice with a
fugacity $\sqrt{Q}$.

\section{Holomorphic observables on the lattice}\label{sectlattice}
\setcounter{equation}{0}

In this Section, we will recall the definition of parafermion (or
spinor) operators, and we will give it an explicit geometrical
meaning in the context of the FK representation. We will then show
that, at a particular value of the spin, the parafermions are
purely holomorphic observables in a lattice sense.

Let us first briefly recall the definitions of order and disorder
operators, which are the basic ingredients in the construction of
lattice parafermions (for more details, see \cite{para,NK}). An
order (or spin) operator $\sigma_i$ is associated to every site of
the original lattice (dots in Fig.\,\ref{FKconfigwired}), taking
values in $\left\{e^{\frac{2\pi i}{Q}\,S_i}\right\}$, where
$S_i\in\{0,1,...,Q-1\}$. The spin--spin correlation function
$\;\langle\sigma_i\sigma_j^*\rangle\,=\,\langle e^{\frac{2\pi
i}{Q}\,(S_i-S_j)}\rangle\;$ is proportional to a sum like
(\ref{FKZ}) where only those graphs $\tilde{G}\subset\{G\}$ appear
on which the sites $i$ and $j$ belong to the same FK cluster. The
contributions of all other terms vanish, since $S_i$ and $S_j$
belong to different clusters and are freely summed over. Disorder
operators $\mu_k$ can be associated to the sites of the dual
lattice (crosses in Fig.\,\ref{FKconfigwired}) by defining their
correlation function as $\;\langle\mu_k\mu_l^*\rangle\,=\,Z'/Z\;$,
where $Z'$ is obtained from (\ref{originalZ}) by modifying
$\delta(S_k,S_l)\to\delta(S_k+1\,(\text{mod}\, Q),S_l)$ along a
path on the dual lattice that connects sites $k$ and $l$. It has
been shown \cite{para,NK} that this definition of the correlation
function does not depend on the particular path chosen, and that
the resulting operator $\mu$ is dual to the spin operator
$\sigma$. Hence the disorder correlation function only gets
contributions from those graphs where the sites belong to the same
cluster on the dual lattice.

If we now look at the covering lattice, which consists of the
sites of both the original and the dual lattice, we can define
parafermion (or spinor) operators $\psi_p(e)$ on each of its edges
$e$ as the product of the spin and the disorder operator at the
sites connected by $e$, times a phase factor which encodes $p$:
\begin{equation}\label{defspinorsigmamu}
\psi_p(e)\,=\,\sigma(e)\,\mu(e)\,e^{-ip\,\theta(\gamma,e)}\;.
\end{equation}
The meaning of the angle $\theta(\gamma,e)$ will be clarified
below. The existence of several spinor operators, labelled by $p$,
at each value of $Q$, is due to the fact that the disorder
operator can be modified by replacing $\delta(S_k,S_l)$ with
$\delta(S_k+n\,(\text{mod}\, Q),S_l)$, with $n=1,2,...,Q$. It was
shown in \cite{NK} that the different parafermions can be labelled
by a fractional index $p$ which has the physical meaning of spin,
since the spinor correlation function acquires a phase factor
$e^{i4\pi p}$ when the spinors circle one another.

In order to give a precise meaning to the angle $\theta(\gamma,e)$
in (\ref{defspinorsigmamu}), we now \em geometrically \em identify
the spinor operators in terms of the fully packed loops in the FK
representation of the Potts model (dashed lines in
Fig.\,\ref{FKconfigwired}). It follows from
(\ref{defspinorsigmamu}) that spinor correlation functions are
observables of such loops, since they only get contributions from
terms in (\ref{FKZ}) where all order and all disorder operators
belong to the same FK cluster on the original and dual lattice,
respectively, hence all edges $e$ belong to the same loop on the
medial lattice. For each term in (\ref{FKZ}) which contributes to
a spinor correlation function, we can define the parafermions'
phases as
\begin{equation}\label{defspinor}
\arg\left[\psi_p(e)\right]\,=\,-p\,\theta(\gamma,e)\;,
\end{equation}
where $\gamma$ indicates the loop which crosses all edges involved
in the given correlation function, and the angle
$\theta(\gamma,e)$ is the inclination at which $\gamma$ intersects
the edge $e$. In the cases of interest $p$ is fractional and
(\ref{defspinor}) defines the left hand side only up to a multiple
of $2\pi p$. However we can give it a unique meaning as follows:
for a curve $\gamma$ which begins and ends on the boundary we
choose $\theta(\gamma,e)$ for $e$ on the boundary to be in the
range $[0,2\pi)$, and then define it recursively at other points
by requiring that between neighbouring edges on the medial lattice
it can change only by $\pm\pi/2$. With this definition we see that
$\theta(\gamma,e)$ can be arbitrarily large, depending on the
degree of winding of $\gamma$ around the edge $e$. If $\gamma$ is
a closed loop, we choose a given edge $e$ at random, require again
that $\theta(\gamma,e)\in[0,2\pi)$, and define the angles of the
other edges recursively as before. Since all observables will turn
out to depend only on the differences of the $\theta(\gamma,e)$
between different edges on the same curve, this arbitrariness is
not important.

\subsection{Lattice Holomorphicity}

We now show that, at the critical point,
$\;\langle\psi_{p}(e_1)\psi_{p}(e_2)\cdots\psi_{p}(e_n)\rangle\;$
is a ``lattice--holomorphic" function of each of its $n$
variables, i.e. it satisfies a discrete version of the
Cauchy--Riemann equations, if we properly choose the value of $p$.
To this purpose, we will prove that, if
\begin{equation}\label{holomorphicspin}
\sqrt{Q}\,=\,2\,\sin\left(p\,\frac{\pi}{2}\right)\;,
\end{equation}
then
\begin{equation}\label{holomorphicity}
\sum\limits_{e\in
C}\,\langle\psi_{p}(e_1)\cdots\psi_{p}(e_n)\,\psi_p(e)\rangle\,\delta
z_e\,=\,0\;,
\end{equation}
for every closed contour $C$ of the covering lattice which does
not include $e_1,...,e_n$\footnote{Singularities occur in the
correlation function when two of its arguments coincide.}. This
equation has a direct meaning as lattice version of a vanishing
contour integral. Furthermore, if we consider $C$ as the boundary
of an elementary plaquette of the covering lattice,
(\ref{holomorphicity}) is equivalent to the discrete
Cauchy--Riemann equations, which constitute its real and imaginary
part (for a systematic study of discrete Cauchy--Riemann equations
see \cite{mercat}).

\begin{figure}[h]
\begin{center}
\psfig{figure=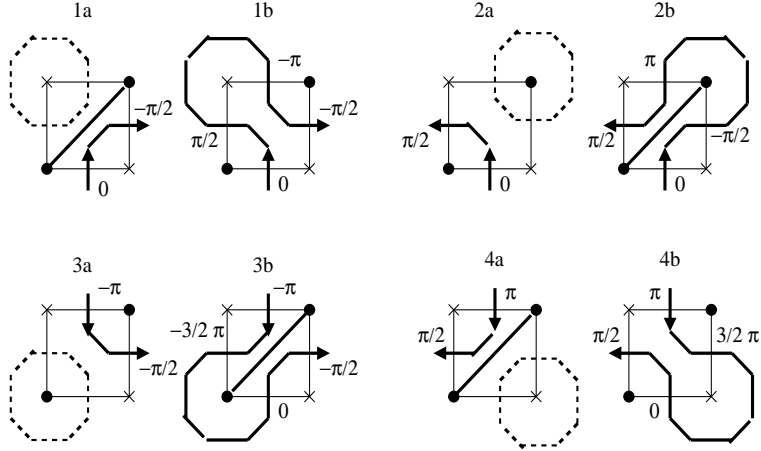,height=6cm,width=10cm}
\end{center}
\caption{Inequivalent configurations on a single
plaquette}\label{classes}
\end{figure}

\textit{Proof of (\ref{holomorphicity})}. Let us first consider
the case when $C$ is the boundary of an elementary plaquette of
the covering lattice. A loop $\gamma$, which intersects all edges
$e_1,...,e_n$ and $e$, can visit the plaquette in eight
inequivalent ways, which are depicted in Fig.\,\ref{classes},
where $\gamma$ is represented by the continuous line\footnote{For
definiteness, we have oriented the loop so the sites of the
original lattice, carrying the Potts spins, lie immediately to its
left, and the dual sites to its immediate right.}. The dashed
lines represent other loops in the same configuration; their paths
outside the plaquette can be different from the picture, but their
particular shape does not matter, as we shall see below. To each
edge it is associated the value of the inclination at which
$\gamma$ intersects the edge in the considered configuration, up
to a multiple of $2\pi$, which is, however, the same for every
edge in the plaquette which the curve visits.

\vspace{0.5cm}

We will now show that configurations $(1a)$ and $(1b)$ cancel each
other in the sum (\ref{holomorphicity}), and the same can be
verified in the other cases. The result automatically holds in all
winding classes, since winding simply implies a global rotation of
all angles.

The sum over (1a) and (1b) gives
\begin{eqnarray}\label{sumab}
&&\left.\sum\limits_{e\in
C}\,\langle\psi_{p}(e_1)\cdots\psi_{p}(e_n)\,\psi_p(e)\rangle\,\delta
z_e\right|_{1a,1b}\,=\,\\\nonumber
&&\,=\,\left(1\,+\,i\,e^{i\,p\,\frac{\pi}{2}}\right)\,P(1a)
\,+\,\left(1\,+\,i\,e^{i\,p\,\frac{\pi}{2}}\,+\,(-1)\,e^{i\,p\,\pi}\,+\,(-i)\,e^{-i\,p\,\frac{\pi}{2}}
\right)\,P(1b)\;,
\end{eqnarray}
where $P(1a)$ and $P(1b)$ are the weights of all graphs in the
class $(1a)$ or $(1b)$, respectively. From (\ref{FPLZ},\ref{uc})
we know that, at criticality,
\begin{equation}
\label{eom} P(1a)\,=\,\sqrt{Q}\,P(1b)\;,
\end{equation}
since the two types of configuration differ from the
presence/absence of a closed loop, weighted with a factor
$\sqrt{Q}$ (independently of its particular shape).
Eq.\,(\ref{eom}) can be interpreted as an "equation of motion",
since it expresses the variation of the partition function under a
local rearrangement of configuration. Therefore, (\ref{sumab})
vanishes for $\sqrt{Q}\,=\,2\,\sin\left(p\,\frac{\pi}{2}\right)$,
i.e. when (\ref{holomorphicspin}) holds. It is easy to check that
the same cancellation mechanism takes place for the other pairs of
configurations, which proves (\ref{holomorphicity}) for the single
plaquette case.

The validity of (\ref{holomorphicity}) for an arbitrary closed
lattice contour $C$ follows directly since it is equal to the sum
of the contour integrals around each elementary plaquette
contained inside $C$.
\begin{flushright}
$\Box$
\end{flushright}

\subsection{One--point function}\label{sect1pt}

\begin{figure}[h]
\begin{center}
\psfig{figure=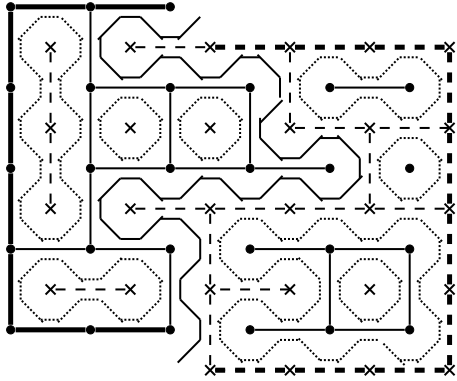,height=5cm,width=6cm}
\end{center}
\caption{A configuration in the FK representation of the Potts
model with wired and free boundary conditions (thick lines on the
original and dual lattice, respectively)}\label{FKconfig}
\end{figure}

By choosing suitable boundary conditions, it is possible to obtain
a non--vanishing expectation value of $\psi_p$. We will now focus
on this situation, in view of its simplicity and its relation to
SLE, which will be discussed in Section\,\ref{sectrelSLElatt}. If
we impose wired boundary conditions on one connected component of
the boundary and free boundary conditions on its complement (see
Fig.\,\ref{FKconfig}), a domain wall $\gamma$ will be generated on
the medial lattice (continuous line), which is simultaneously the
boundary of the cluster attached to the wired part of the
boundary, and of the dual cluster attached to the rest of the
boundary. In addition, of course, there will in general be closed
loops (dashed lines).

\vspace{0.5cm}

The expectation value $\langle\psi_p(e)\rangle$ vanishes every
time the values of $\sigma(e)$ and/or $\mu(e)$ can be summed over
freely in the partition function. As a consequence,
$\langle\psi_p(e)\rangle$ is different from zero only on the edges
crossed by the domain wall, since in this case both $\sigma(e)$
and $\mu(e)$ belong to a cluster connected to the boundary and
therefore have a fixed value (wired boundary conditions are
equivalent to fixing the value of $\sigma$ and free boundary
conditions to fixing $\mu$). Therefore, we can define the phase
$\theta(\gamma,e)$ in (\ref{defspinor}) as the inclination at
which the domain wall $\gamma$ intersects the edge $e$. It is easy
to check that the proof of (\ref{holomorphicity}) also holds for
\begin{equation}\label{holomorphicity1pt}
\sum\limits_{e\in C}\,\langle\psi_p(e)\rangle\,\delta z_e\,=\,0\;,
\end{equation}
by simply considering $\gamma$ as the domain wall described above.
Therefore, $\langle\psi_p(e)\rangle$ is a lattice holomorphic
function.

Moreover, we can determine which are the boundary conditions
satisfied by $\langle\psi_p(e)\rangle$. To this purpose, we have
to look at the plaquettes of the covering lattice which are cut in
half by the boundary edges of the original or dual lattice, and
therefore have a triangular shape. The presence of the boundary
implies that $\langle\psi_p(e)\rangle$ only differs by a phase on
the two edges of any of these plaquettes, since the domain wall
automatically intersects the second edge of a plaquette once it
entered the first. Therefore, we can limit our analysis to half of
these edges, and we choose for convenience the entering edges
$e_L$ of the plaquettes which lie on the left side of the boundary
(with wired boundary conditions), and the exiting edges $e_R$ of
the plaquettes which lie on the right side (with free boundary
conditions). Since winding is not possible around these
plaquettes, the phase $\theta(\gamma,e)$ is uniquely determined by
the boundary edge under consideration:
$$
\theta(e_L)=\theta_b-\pi\;,\qquad\qquad \theta(e_R)=\theta_b\;,
$$
where $\theta_b$ is the inclination of the boundary itself
(measured counterclockwise from the starting point of the curve)
and we set to zero the value of $\theta(\gamma,e)$ at the starting
edge in Fig.\,\ref{FKconfig} (these arguments can be easily
generalized to more general domains than the one in
Fig.\,\ref{FKconfig}). Therefore, the phase of
$\langle\psi_p(e)\rangle$ on the boundary is fixed as
\begin{equation}\label{latticebc}
\text{arg}[\langle\psi_p(e_L)\rangle]=-p\,(\theta_b-\pi)\;,\qquad\qquad
\text{arg}[\langle\psi_p(e_R)\rangle]=-p\,\theta_b\;.
\end{equation}
This is the generalization to an arbitrary domain of the boundary
conditions satisfied by the function $z^{-p}$ on the upper half
plane when the domain wall starts from $x=0$ on the real axis,
since in that geometry our convention implies $\theta_b=0$ for
$x>0$ and $\theta_b=2\pi$ for $x<0$.

\section{The Ising model}\label{sectising}
\setcounter{equation}{0}

In this Section, we will analyse in further detail the properties
of $\langle\psi_p(e)\rangle$, enlightening some special features
of the case $Q=2$, i.e. the Ising model (for previous studies of
the fermion operator in this particular case, see
\cite{para,mercat,DD}).

\subsection{Additional equation}

Let us start by noticing that, in general,
eq.\,(\ref{holomorphicity1pt}) is not sufficient to determine the
function $\langle\psi_p(e)\rangle$. The reason is that, on a
covering lattice with $N$ sites, we have to determine the complex
function $\langle\psi_p(e)\rangle$ on $2N$ edges, but we only have
the single complex equation (\ref{holomorphicity1pt}) for each of
the $O(N)$ plaquettes.

However, in the particular case $Q=2$ and $p=\frac{1}{2}$ it is
possible to see that another complex relation holds for every
elementary plaquette. This has the form
$$
\langle\psi_{1/2}(e_1)\rangle-\langle\psi_{1/2}(e_2)\rangle+\langle\psi_{1/2}(e_3)\rangle
-\langle\psi_{1/2}(e_4)\rangle\,=\,0\;,
$$
where $e_1,...,e_4$ are the four consecutive edges of the
plaquette.

In order to formulate this in a more systematic fashion, let us
consider again the eight classes of configuration in
Fig.\,\ref{classes}. Let us define
\begin{equation}\label{defgi}
g_i\,=\,\sum\limits_{w=-\infty}^{\infty}\sum\limits_{G\in
C_{ib}^{(w)}}e^{-ip\,2\pi w}\,P(G)\;,
\end{equation}
where $C_{ib}^{(w)}$ is the set of all graphs in the FK expansion
which produce configurations of type $(ib)$ $(i=1,..,4)$, with
winding number $w$, and $P(G)$ is the weight of a graph $G$. It
follows from (\ref{FPLZ}) that for every $G\in C_{ib}$ there is a
corresponding $G\in C_{ia}$ such that
$$
P(G_{1a})=u\,P(G_{1b})\;, \qquad
P(G_{2a})=\frac{Q}{u}\,P(G_{2b})\;,\qquad
P(G_{3a})=\frac{Q}{u}\,P(G_{3b})\;,\qquad
P(G_{4a})=u\,P(G_{4b})\;.
$$
(at criticality all factors reduce to $\sqrt{Q}$). We therefore
have
\begin{eqnarray}\label{psiei}
\langle\psi_p(e_1)\rangle&=&\left\{(u+1)\,g_1\,+\,\left(\frac{Q}{u}+1\right)\,g_2
\,+\,g_3\,+\,g_4\right\}\\\nonumber
\langle\psi_p(e_2)\rangle&=&\left\{(u+1)\,e^{ip\frac{\pi}{2}}\,g_1
\,+\,e^{ip\frac{\pi}{2}}\,g_2\,+\,
\left(\frac{Q}{u}+1\right)\,e^{ip\frac{\pi}{2}}\,g_3\,+\,e^{-ip\frac{3}{2}\pi}\,g_4\right\}\\\nonumber
\langle\psi_p(e_3)\rangle&=&\left\{e^{ip\pi}\,g_1\,+\,e^{-ip\pi}\,g_2\,+\,
\left(\frac{Q}{u}+1\right)\,e^{ip\pi}\,g_3\,+\,(u+1)\,e^{-ip\pi}\,g_4\right\}\\\nonumber
\langle\psi_p(e_4)\rangle&=&\left\{e^{-ip\frac{\pi}{2}}\,g_1
\,+\,\left(\frac{Q}{u}+1\right)\,e^{-ip\frac{\pi}{2}}
\,g_2\,+\,e^{ip\frac{3}{2}\pi}\,g_3\,+\,(u+1)\,e^{-ip\frac{\pi}{2}}\,g_4\right\}
\end{eqnarray}
which can be written in the matrix notation
\begin{equation}\label{matrix}
\langle\psi\rangle\,=\,M \cdot g\;.
\end{equation}
It is now easy to check that
\begin{equation}\label{rank}
\text{rank} (M)\,=\,\begin{cases}\;4\quad\text{for generic
}\sqrt{Q}\\\;3\quad\text{for }\sqrt{Q}=2\sin
p\frac{\pi}{2}\\\;2\quad\text{for
}Q=2\,,\;p=\frac{1}{2}\end{cases}\qquad \forall u\;.
\end{equation}
Thus in general, if we choose $p$ correctly, there is one linear
relation between the parafermions around the 4 edges of each
plaquette, while for $Q=2$ there are two.

\subsection{Continuum limit of the lattice equations}
In this section we make some remarks on the question of the extent
to which our results on lattice holomorphicity imply that in the
continuum limit the parafermion correlators are truly holomorphic
functions of their arguments. In general this does not follow
without further assumptions. Recall Morera's theorem\cite{Morera}
which states that if $\oint_Cf(z)dz$ vanishes for every closed
contour $C$ in some domain $\cal D$, \em and \em $f(z)$ is
continuous, then $f(z)$ is a complex analytic function in $\cal
D$. Although our lattice result shows that there is a subsequence
of approximations to $\oint_Cf(z)dz$ which do vanish identically,
this does not imply that the limit vanishes without further
smoothness assumptions. Although we expect these to be valid, it
would require further work to show that they follow from the
lattice model. Nevertheless, if true, this implies that in, for
example, the upper half plane when the curve runs from $0$ to
infinity, $\langle\psi_p(z)\rangle\to{\rm const.}\,z^{-p}$.

In the case of the Ising model, $Q=2$, the situation is better. In
that case there are sufficient equations to determine the lattice
function $\langle\psi(e)\rangle$. Since they are linear and
translationally invariant, it should be a simple matter to solve
then by Fourier analysis in a sufficiently simple domain, and then
to show that, in the limit when the size of the domain is much
larger than the lattice spacing, the Fourier components near $k=0$
dominate, and in the limit, are given by the solution of the
continuum problem. However, this approach is not adequate for more
irregular domains.

\subsection{Off--criticality}

As soon as we move away from the critical point,
eq.\,(\ref{holomorphicity1pt}) does not hold anymore, since the
necessary cancellation mechanism relies on the particular value
$u_c=\sqrt{Q}$. However, linear relations between the values of
$\langle\psi_p\rangle$ at different edges are still present if
$\sqrt{Q}=2\sin p\frac{\pi}{2}$, since (\ref{rank}) holds for any
value of $u$. For instance, if we define $v=u-\sqrt{Q}$ we have
\begin{equation}\label{offcritgeneral}
\langle\psi_p(e_1)\rangle+i\langle\psi_p(e_2)\rangle-\langle\psi_p(e_3)\rangle-i\langle\psi_p(e_4)\rangle
\,=\,X\,(\langle\psi_p(e_3)\rangle-\langle\psi_p(e_1)\rangle)\;,
\end{equation}
where
$$
X\,=\,v\;\frac{1+ie^{ip\frac{\pi}{2}}}{e^{ip\pi}-1-2\sin
p\frac{\pi}{2}-v}\;.
$$

We will now give a precise meaning to this off--critical relation
in the case $Q=2$. It is well known that the Ising model can be
described, in the scaling regime near its phase transition, by a
field theory of a free Majorana fermion of chiral components
$\chi,\bar{\chi}$, with the action
\begin{equation}\label{actionmajorana}
S\,=\,\frac{1}{2\pi}\,\int\,d^2z\,\left(\chi\partial_{\bar{z}}\chi\,+\,
\bar{\chi}\partial_{z}\bar{\chi} \,+\,im\bar{\chi}\chi\right)\;,
\end{equation}
where the parameter $m$ measures the deviation from the critical
temperature. In the above notation, $m$ is positive for $T<T_c$
and negative for $T>T_c$. The absolute value $|m|$ is the mass of
the elementary excitation in both regimes. The action
(\ref{actionmajorana}) implies the equations of motion
\begin{equation}\label{eomCFT}
\partial_{\bar{z}}\chi=i\frac{m}{2}\,\bar{\chi}\;,
\qquad\qquad \partial_{z}\bar{\chi}=-i\frac{m}{2}\,\chi\;.
\end{equation}

The left hand side of (\ref{offcritgeneral}) is the discrete
version of $2i\partial_{\bar{z}}\langle\psi_p\rangle$, therefore
we have to reexpress the right hand side in terms of
$\langle\bar{\psi}_p\rangle$ in order to reproduce a discrete
version of (\ref{eomCFT}). This is possible only in the case
$Q=2$, because at the value $p=\frac{1}{2}$ the functions $g_i$
defined in (\ref{defgi}) coincide with their complex conjugates
$\bar{g}_i$. In more general cases, the right hand side of
(\ref{offcritgeneral}) cannot be just written in terms of
$\langle\bar{\psi}_p\rangle$, and this is consistent with the fact
that no simple equation like (\ref{eomCFT}) holds in the
corresponding field theory descriptions. Going back to the Ising
case, we have
\begin{eqnarray*}
&&\langle\psi_{1/2}(e_1)\rangle+i\langle\psi_{1/2}(e_2)\rangle-\langle\psi_{1/2}(e_3)\rangle-i\langle\psi_{1/2}(e_4)\rangle
\,=\,\\
&&\hspace{2cm}\,=\,i\frac{v}{(1+\sqrt{2})(2\sqrt{2}+v)}\,\left(\langle\bar{\psi}_{1/2}(e_1)\rangle+\langle\bar{\psi}_{1/2}(e_2)\rangle
+\langle\bar{\psi}_{1/2}(e_3)\rangle+\langle\bar{\psi}_{1/2}(e_4)\rangle\right)\;,
\end{eqnarray*}
which is the discrete version of
$$
2i\partial_{\bar{z}}\langle\psi_{1/2}\rangle\,=\,i\frac{v}{(1+\sqrt{2})(2\sqrt{2}+v)}\,4\langle\bar{\psi}_{1/2}\rangle
$$
where the antiholomorphic operator in the right hand side is
uniformly distributed on the plaquette. This relation and its
complex conjugate are equivalent to (\ref{eomCFT}) if the fields
are related by
$$
\chi\,=\,\alpha\,\psi_{1/2}\;,\qquad
\bar{\chi}\,=\,\beta\,\bar{\psi}_{1/2}\;, \qquad\text{with}\quad
\frac{\alpha}{\beta}=i
$$
and the parameter $m$ is given by
\begin{equation}\label{mlattice}
m=4\frac{v}{(1+\sqrt{2})(2\sqrt{2}+v)}=4\frac{e^J-1-\sqrt{2}}{(1+\sqrt{2})(\sqrt{2}-1+e^J)}
\,\simeq\,\sqrt{2}\,(J-J_c)+O(J-J_c)^2\;.
\end{equation}
We can now compare this result with the known value of the
correlation length $\xi$ on the square lattice, expressed in terms
of the modular parameter $k=(\sinh J)^{-2}$ \cite{baxter} and
evaluated in the scaling limit:
\begin{equation}\label{xilattice}
\frac{1}{\xi}\,=\,\begin{cases}\;-\log
k\,\simeq\,2\sqrt{2}\,(J-J_c)+O(J-J_c)^2\qquad\;\text{for}\quad
T<T_c\\\;\frac{1}{2}\log
k\,\simeq\,\sqrt{2}\,(J_c-J)+O(J_c-J)^2\qquad\quad\text{for}\quad
T>T_c\end{cases}
\end{equation}
From the definition of the correlation length in terms of the
connected two--point function of the order operator $\sigma$
$$
\langle\sigma(x)\sigma(0)\rangle\,\sim\,x^{-\tau}\,e^{-x/\xi}\qquad\text{as}\quad
x\to\infty \;,
$$
it follows that $1/\xi$ must be equal to the mass of the lightest
excitation which couples to $\sigma$. The $\mathbb{Z}_2$ symmetry
of the Ising model implies that $\sigma$ only couples to states
with odd numbers of particles at high temperature and with even
number of particles at low temperature. Hence (\ref{mlattice}) and
(\ref{xilattice}) are in perfect agreement.

\section{Continuum predictions}\label{sectcontlim}
\setcounter{equation}{0}

In this Section, we will show that, despite difficulties in
proving convergence to the continuum limit, the lattice
holomorphic objects defined above nevertheless have their
counterparts in the continuum formulations of these models. In
particular, we will review results obtained by CFT and Coulomb gas
methods, and we will describe a new prediction from SLE.

\subsection{CFT and Coulomb gas results}\label{sectcoulomb}

From the CFT point of view, the holomorphicity condition
(\ref{holomorphicspin}) has a clear meaning both for $Q=2$ and
$Q=3$, where the values $p=\frac{1}{2}$ and $p=\frac{2}{3}$
coincide with the spins of the parafermionic currents predicted by
Fateev and Zamolodchikov \cite{zamfat} for CFT with extended
$\mathbb{Z}_2$ and $\mathbb{Z}_3$ symmetry, respectively.

A more general result was obtained in \cite{NK} for the Potts
model by using Coulomb gas methods. Nienhuis and Knops showed that
for every $Q$, there is a series of spinor operators parameterized
by a rational number $p<1$, whose denominator is the order of one
of the cyclic permutations into which the permutations of
$\{0,1,...,Q-1\}$ can be decomposed. For instance, at $Q=2$ we
only have $p=1/2$, while at $Q=3$ we have $p=1/2$, $1/3$ or $2/3$.
The scaling dimension $x_p$ and spin $s_p$ of the spinors are
given by
\begin{equation}\label{nienknops}
x_p\,=\,1\,+\,\frac{1}{2-y}\,(p^2-1)\;,\qquad\qquad s_p\,=\,p\;,
\end{equation}
where $y$ is defined through
\begin{equation}
\sqrt{Q}\,=\,2\,\cos\left(\pi\, \frac{y}{2}\right)\;.
\end{equation}
For every $Q$, there is a single holomorphic operator among these,
characterized by $x_{p^*}=s_{p^*}$, i.e.
\begin{equation}\label{nienknopshol}
p^*\,=\,1-y\;.
\end{equation}
This result is in perfect agreement with our requirement
(\ref{holomorphicspin}) for the holomorphicity of the
parafermionic correlation functions on the lattice.

\subsection{The SLE prediction}\label{sectSLE}

In this Section, we will show that some local observables
associated to an SLE curve are characterized by the scaling
dimension of the spinor operators discussed above. To this
purpose, we will recall and generalize results obtained in
\cite{Tbulk} for the analysis of another operator, the
stress--energy tensor.

Let us consider a chordal SLE$_\kappa$ process (for $0<\kappa<8$)
on the upper half plane described by complex coordinates
$w,\bar{w}$ \cite{schramm}. The probability
$P(w_1,w_2,\bar{w}_1,\bar{w}_2)$ that the curve passes between two
points $w_1,\,w_2$ (or to the left or right of both) satisfies the
equation
\begin{equation}\label{eq12}
\left\{\frac{\kappa}{2}\left(\partial_{w_1}+\partial_{\bar{w}_1}+\partial_{w_2}+\partial_{\bar{w}_2}\right)^2
+\frac{2}{w_1}\partial_{w_1}+\frac{2}{\bar{w}_1}\partial_{\bar{w}_1}+\frac{2}{w_2}\partial_{w_2}
+\frac{2}{\bar{w}_2}\partial_{\bar{w}_2}\right\}P(w_1,w_2,\bar{w}_1,\bar{w}_2)\,=\,0\;.
\end{equation}
If we parameterize the event by the middle point $w$ of a straight
segment, by its length $\epsilon$ and by the angle $\theta$ that
it makes with the positive imaginary direction:
$$
w_1=w-\frac{\epsilon}{2}e^{i\theta}\;,\qquad
w_2=w+\frac{\epsilon}{2}e^{i\theta}\;,
$$
we obtain, at leading order in $\epsilon$, the following equation
\begin{equation}\label{eqP}
\left\{\frac{\kappa}{2}\left(\partial_{w}+\partial_{\bar{w}}\right)^2
+\frac{2}{w}\partial_{w} +\frac{2}{\bar{w}}
\partial_{\bar{w}}-\left(\frac{1}{w^2}+\frac{1}{\bar{w}^2}\right)\epsilon\partial_{\epsilon}+
\left(\frac{1}{w^2}-\frac{1}{\bar{w}^2}\right)i\partial_{\theta}+O(\epsilon^2)\right\}P(w,\bar{w},
\epsilon,\theta)\,=\,0\;.
\end{equation}
In the above expansion we have assumed that each of the Fourier
modes of the probability, defined as
\begin{equation}\label{prelimdefQn}
Q_{p}(w,\bar{w},\epsilon)=\int\limits_{-\infty}^{\infty} d\theta
\,e^{-ip\,\theta}\,P(w,\bar{w},\epsilon,\theta)\;,
\end{equation}
vanishes with a power law $\epsilon^{x_p}$ as $\epsilon\to 0$,
hence $\partial_{\epsilon}=O(\epsilon^{-1})$. The parameter $p$
has the physical meaning of "spin" and took the value $p=2$ in the
case of the stress-energy tensor studied in \cite{Tbulk}. Here we
are interested in more general values of $p$, in particular
non-integer values. Therefore, the domain of integration for the
variable $\theta$ in (\ref{prelimdefQn}) is not restricted between
$0$ and $2\pi$, since the winding of the SLE curve around the
segment is associated to a shift $\theta+2\pi n$, where
$n\in\mathbb{Z}$ is the winding number. If $p$ is integer this
phenomenon is irrelevant, but if $p<1$ winding has to be taken
into account. In particular, there are $d$ inequivalent winding
classes, where $d$ is the denominator of $p$. We can define the
probability that the SLE curve passes between the ending points of
the segment with a winding number $n=k\;(\text{mod}\;d)$ as
$$
P_k(w,\bar{w},\epsilon,\theta)\,=\,\sum\limits_{m=-\infty}^{\infty}P(w,\bar{w},\epsilon,\theta+2\pi
k+2\pi m d)\;.
$$
In this notation, the Fourier modes can be expressed as
\begin{equation}\label{defQn}
Q_{p}(w,\bar{w},\epsilon)\,=\,\sum\limits_{k=0}^{d-1}e^{-ip\,2\pi
k}\int\limits_{0}^{2\pi} d\theta
\,e^{-ip\,\theta}\,P_k(w,\bar{w},\epsilon,\theta)\;.
\end{equation}
By integrating eq.\,(\ref{eqP}) over $\int\limits_{0}^{2\pi}
d\theta \,e^{-ip\,\theta}$ we obtain, to leading order in
$\epsilon$,
\begin{equation}\label{eqQn}
\left\{\frac{\kappa}{2}(\partial_{w}+\partial_{\bar{w}})^2+\frac{2}{w}\,\partial_{w}
+\frac{2}{\bar{w}}\,\partial_{\bar{w}}
-\left(\frac{1}{w^2}+\frac{1}{\bar{w}^2}\right)\,\epsilon\,\partial_{\epsilon}
-p\,\left(\frac{1}{w^2}-\frac{1}{\bar{w}^2}\right)\right\}\,Q_p(w,\bar{w},
\epsilon)+\text{corrections}\,=\,0\;,
\end{equation}
Among the several solutions of (\ref{eqQn}), we are now interested
in the one which satisfies appropriate boundary conditions for the
event of passing in between the two initial points. This
particular solution was obtained in \cite{Tbulk} by mapping the
problem into the disk geometry; for completeness, we report the
detailed discussion in Appendix\,\ref{appdisk}. The solution is
\begin{equation}\label{Qnsol}
Q_p(w,\bar{w}, \epsilon)\,=\,c_p
\,\epsilon^{\,x_p}\,w^{\alpha_p}\,\bar{w}^{\beta_p}\,(w-\bar{w})^{\gamma_p}\;,
\end{equation}
with
\begin{equation}\label{xn}
\alpha_p=\frac{\kappa-8}{2\kappa}-\frac{p}{2}\;,\quad
\beta_p=\frac{\kappa-8}{2\kappa}+\frac{p}{2}\;,\quad
\gamma_p=\frac{(8-\kappa)^2-\kappa^2p^2}{8\kappa}\;,\quad
x_p=1-\frac{\kappa}{8}+\frac{\kappa}{8}\,p^2\;.
\end{equation}
The function (\ref{Qnsol}) gives the correct leading behaviour for
small $\epsilon$ of $Q_{p}(w,\bar{w},\epsilon)$ only when the
higher order terms in (\ref{eqP}) do not mix the Fourier
components at leading order. This is guaranteed if
\begin{equation}\label{condition}
x_p\,<\,2m+x_{p-2m} \qquad \qquad \forall\; m=1,2,3,...\quad
\text{such that}\quad c_{p-2m}\neq 0\;,
\end{equation}
where $m$ labels the powers of $\epsilon^2$ in the expansion of
the differential equation (\ref{eq12}). This automatically holds
for any $0\leq p\leq 1$, which is the case of interest in this
paper.

At the particular value
\begin{equation}\label{kappaofn}
\kappa_{p}=\frac{8}{p+1}\;,
\end{equation}
(\ref{Qnsol}) simplifies to the purely holomorphic function
\begin{equation}\label{Qnhol}
Q_{p}(w,\bar{w},\epsilon)=\text{const}\,\times\,\left(\frac{\epsilon}{w}\right)^p\;,
\end{equation}
where spin and scaling dimension are equal.

In order to compare the SLE prediction with the other results
described in this paper, let us recall that the $Q$--states Potts
model in the FK representation is conjectured \cite{basicSLE} to
be described, in the continuum limit, by SLE$_{\kappa}$ where
\begin{equation}\label{conjSLEPotts}
\sqrt{Q}\,=\,-2\cos\left(\pi\,\frac{4}{\kappa}\right)\;,\qquad
4<\kappa<8\;.
\end{equation}
It is then easy to check that (\ref{xn}) perfectly agrees with the
Coulomb gas results (\ref{nienknops}) for every $0<p<1$, and in
particular at the holomorphic value (\ref{kappaofn}).

The holomorphic solution (\ref{Qnhol}) has a natural CFT
interpretation as the expectation value of a purely holomorphic
operator of spin $p$, given the appropriate boundary conditions
which generate the domain wall. By recalling the relation between
$\kappa$ and the central charge $c$ of CFT \cite{BB}
\begin{equation}\label{cofkappa}
c(\kappa)=\frac{(3\kappa-8)(6-\kappa)}{2\kappa}\;,
\end{equation}
one can see that (\ref{kappaofn}) implies
\begin{equation}\label{nCFT}
p\,=\,\begin{cases}\;h_{3,1}\qquad\text{for}\quad 0<\kappa<4\\
\;h_{1,3}\qquad\text{for}\quad 4<\kappa<8\end{cases}\;,
\end{equation}
where $h_{i,j}$ are the conformal weights of primary operators in
the Kac table of CFT. For Ising and 3--state Potts models, this is
again in agreement with the other results presented in this paper.
Relation (\ref{nCFT}), however, is more general and deserves
further study in connection with other statistical models.

\section{Relation between SLE and lattice results}\label{sectrelSLElatt}
\setcounter{equation}{0}

In the last Section, we have checked that the SLE prediction
agrees with other continuum limit results. Here, we will discuss
the explicit relation between the purely holomorphic function
(\ref{Qnhol}) obtained from SLE and the lattice spinor $\psi_p$
defined in (\ref{defspinor}).

To this purpose, we have to look at the one--point function
$\langle\psi_p(e)\rangle$, which is an observable of the domain
wall that is conjectured to converge to SLE in the continuum
limit. We have shown in Sect.\,\ref{sect1pt} that
$\langle\psi_p(e)\rangle$ is purely holomorphic when
$\sqrt{Q}=2\sin p\,\frac{\pi}{2}$, i.e. when
$\kappa=\frac{8}{p+1}$ as in (\ref{kappaofn}). Moreover, we know
that $\langle\psi_p(e)\rangle$ satisfies the same boundary
conditions obeyed by $z^{-p}$ in the continuum limit, hence we
expect
\begin{equation}\label{Qproppsi}
Q_{p}(w,\bar{w},\epsilon)=\text{const}\,\times\,\epsilon^p\,\langle\psi_p(w)\rangle^{\text{cont}}\;,
\end{equation}
where $\langle\psi_p(w)\rangle^{\text{cont}}$ is the function to
which $\langle\psi_p(e)\rangle$ is supposed to converge in the
continuum limit\footnote{For $p\neq 1/2$, we need a continuity
assumption to justify this statement, as discussed in
Sect.\,\ref{sectising}.}.

\begin{figure}[h]
\begin{center}
\psfig{figure=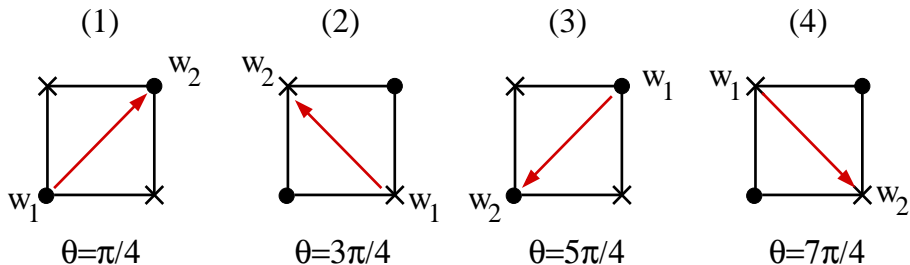,height=3.5cm,width=12cm}
\end{center}
\caption{Rotating segment within a single
plaquette.}\label{slitpaper}
\end{figure}

The advantage of SLE, with respect to other continuum limit
approaches, is to have a direct and explicit meaning in terms of
the lattice formulation of the problem. This allows us to analyse
the correspondence between $Q_p(w,\bar{w},\epsilon)$ and
$\langle\psi_p(e)\rangle$ on the lattice itself, by defining a
discretized version $Q_p^{\text{lat}}$ of $Q_p$ where the SLE
probability distributions is replaced by the lattice weights in
the FK representation of the Potts model. In principle, one should
check the relation for segments connecting two arbitrary points
$w_1,w_2$ on the lattice. As a matter of fact, the computation
becomes very complicated as soon as several plaquettes are
involved, and we have been able to carry it out explicitly only
for the case of a single plaquette. This corresponds to the
rotating segment shown in Fig.\,\ref{slitpaper}.

Let us first obtain the lattice version of $Q_p$. The range of
integration in (\ref{defQn}) can be further reduced to
$\theta\in[0,\pi]$ by noticing that, in every winding sector, the
curve can pass between the end points of the segment by leaving
point $w_1$ to its left and $w_2$ to its right, or viceversa, and
the two events are related by the transformation
$\theta\to\theta+\pi$:
$$
Q_p(w,\bar{w},\epsilon)\,=\,\sum_k\,e^{-ip\,2\pi
k}\,\int\limits_0^{\pi}d\theta\,e^{-ip\,\theta}\,\left[P_{k,1L,2R}(w,\bar{w},\epsilon,\theta)
+e^{-ip\,\pi}P_{k,1R,2L}(w,\bar{w},\epsilon,\theta+\pi)\right]\;.
$$
This implies that we only have to analyse cases $(1)$ and $(2)$ in
Fig.\,\ref{slitpaper}. The discretisation of the integral in
$\theta$ gives
\begin{eqnarray}\nonumber
Q^{\text{lat}}_p&=&\frac{\pi}{2}\,\sum_k\,e^{-ip\,2\pi
k}\,\left\{e^{-ip\,\frac{\pi}{4}}\,\left[P^{(1)}_{k,1L,2R}
+e^{-ip\,\pi}P^{(1)}_{k,1R,2L}\right]+\,e^{-ip\,\frac{3}{4}\pi}\,\left[P^{(2)}_{k,1L,2R}
+e^{-ip\,\pi}P^{(2)}_{k,1R,2L}\right]\right\}\,=\\
&=&\frac{\pi}{2}\,\left\{e^{-ip\,\frac{\pi}{4}}\,\left[\sqrt{Q}\,g_2
+e^{-ip\,\pi}\,e^{ip\,2\pi}\sqrt{Q}\,g_3\right]+\,e^{-ip\,\frac{3}{4}\pi}\,\left[\sqrt{Q}\,g_4
+e^{-ip\,\pi}\,e^{ip\,2\pi}\,\sqrt{Q}\,g_1\right]\right\}
\label{Qnlat}
\end{eqnarray}
where the functions $g_i$ are defined in (\ref{defgi}), and their
relation to the probabilities of passing between the end points of
the segment can be inferred from Fig.\,\ref{classes}. The factors
$e^{ip\,2\pi}$ in the last line are due to the fact that the
domain wall in configurations $(1a)$ and $(3a)$ of
Fig.\,\ref{classes} passes between the end points of the segment
with winding $-1$.

We expect from (\ref{Qproppsi}) that (\ref{Qnlat}) is proportional
to $\langle\psi_p(w)\rangle$ evaluated at the middle point $w$ of
the segment, i.e. at the center of the plaquette. Assuming the
spinor operator to be uniformly distributed on the plaquette, we
have
$$
\langle\psi_p(w)\rangle\,=\,\frac{1}{4}\left(\langle\psi_p(e_1)\rangle+\langle\psi_p(e_2)\rangle+
\langle\psi_p(e_3)\rangle+\langle\psi_p(e_4)\rangle\right)\,=\,\frac{1}{4}\sum\limits_{i=1}^{4}S_i\,g_i\;,
$$
where the coefficients $S_i$ can be easily obtained from
(\ref{psiei}) at $u=\sqrt{Q}$. It is now easy to check that
\begin{equation}\label{Qproppsilat}
Q_p^{\text{lat}}\,=\,2\pi\,\frac{\sqrt{Q}\,\sec
p\,\frac{\pi}{4}}{2\left[\sqrt{Q}+2\cos
p\,\frac{\pi}{2}\right]}\,\langle\psi_p(w)\rangle\;,
\end{equation}
which confirms (\ref{Qproppsi}).

It is worth noting that (\ref{Qproppsilat}) is valid for any value
of $Q$, not only for the purely holomorphic case $\sqrt{Q}=2\sin
p\,\frac{\pi}{2}$. This agrees with the general SLE prediction
(\ref{Qnsol}) that, for every $Q$, the several spinor operators
obtained by Coulomb gas methods in \cite{NK} are local observables
of the domain wall.

Let us conclude this Section by analysing the slightly different
situation in which ${\cal Q}_p$ is defined as the Fourier
component of the probability that the domain wall
\textit{intersects} the segment without necessarily passing trough
its end points. Additional configurations have to be taken into
account (see Fig.\,\ref{classes}), and (\ref{Qnlat}) is modified
as
\begin{equation}
{\cal
Q}^{\text{lat}}_p\,=\,Q^{\text{lat}}_p\,+\,\frac{\pi}{2}\,\left\{e^{-ip\,\frac{\pi}{4}}\,\left[g_1
+e^{-ip\,\pi}\,g_4\right]+\,e^{-ip\,\frac{3}{4}\pi}\,\left[g_2
+e^{-ip\,\pi}\,e^{ip\,2\pi}\,g_3\right]\right\}
\;,\label{Qnlatintersect}
\end{equation}
which still implies a proportionality with
$\langle\psi_p(w)\rangle$:
\begin{equation}
{\cal
Q}_p^{\text{lat}}\,=\,2\pi\,\frac{\left(\sqrt{Q}+e^{-ip\,\frac{\pi}{2}}\right)\,\sec
p\,\frac{\pi}{4}}{2\left[\sqrt{Q}+2\cos
p\,\frac{\pi}{2}\right]}\,\langle\psi_p(w)\rangle\;.
\end{equation}
This is consistent with the SLE result, because the boundary
conditions that select solution (\ref{Qnsol}) are the same if we
consider the events of intersecting the segment or passing between
its end points (in the disk geometry defined in
Appendix\,\ref{appdisk}, both events require the curve to pass by
the center of the disk). Therefore, if we assume that the
distortion of the segment under the Loewner conformal map only
introduces terms of higher order in $\epsilon$ in (\ref{eqP}) and
(\ref{eqQn}), $Q_p$ and ${\cal Q}_p$ are expected to only differ
by a constant in the continuum limit, and this is nicely confirmed
by the lattice example analysed here.

\section{Conclusions}\label{sectconclusions}
\setcounter{equation}{0}

In this paper, we have carried out a detailed study of
parafermionic operators in the $Q$--state Potts model, both on the
lattice and in the continuum limit.

Our discrete setting was a planar square lattice. By geometrically
defining the parafermions in terms of the FK representation of the
model, we obtained an explicit proof of lattice holomorphicity for
certain values of the spin. We believe that this is first time
that these parafermions have been identified within the FK
representation of the Ising and Potts models, and that their
holomorphicity has been directly demonstrated at this level. It
would be interesting to extend our analysis to different lattices
and geometries.

After reviewing known results on the scaling dimensions of
parafermions from CFT and Coulomb gas methods in the continuum
limit, we have shown that the same scaling dimensions are
associated to local observables of SLE$_{\kappa}$ when
(\ref{conjSLEPotts}) holds, i.e. at the value of $\kappa$ which is
conjectured to describe the $Q$--state Potts model in the FK
representation. Moreover, we have explicitly discussed the
relation between the SLE observable and the lattice parafermions,
both in the discrete and continuum settings.

Natural extensions of this work are the full analysis of the
situation when the parafermionic operators are not purely
holomorphic, and the study of the SLE prediction in the context of
other statistical systems. In particular, it will be interesting
to exploit the information provided by SLE for models associated
with non--unitary or non--minimal CFT.

\section*{Acknowledgments}

After we had established the lattice holomorphicity equations of
this paper we learned that they had also been derived (together
with similar equations for the O$(n)$ model on a hexagonal
lattice) by S.~Smirnov in the course of his programme to show
convergence of lattice curves to SLE \cite{SmirnovICM}. One of us
(JLC) thanks him for enlightening discussions on this subject.

We also thank B. Doyon and A. Sportiello for useful discussions.
This work was supported by EPSRC under the grant GR/R83712/01. It
was completed while JLC was a visitor to the Kavli Institute for
Theoretical Physics, Santa Barbara, supported by the NSF under
Grant No. PHY99-07949.

\begin{appendix}

\section{SLE probabilities in the disk geometry}\label{appdisk}\setcounter{equation}{0}

The \textit{ansatz}
\begin{equation}\label{Qnsolapp}
Q_p(w,\bar{w}, \epsilon)\,=\,c_p
\,\epsilon^{\,x_p}\,w^{\alpha_p}\,\bar{w}^{\beta_p}\,(w-\bar{w})^{\gamma_p}\;
\end{equation}
solves eq.\,(\ref{eqQn}) for two different choices of the
parameters:
\begin{equation}\label{xndiscarded}
\alpha_p=-\frac{2p}{\kappa-4}\;,\quad
\beta_p=\frac{2p}{\kappa-4}\;,\quad \gamma_p=-\frac{2\kappa
p^2}{(\kappa-4)^2}\;,\quad x_p=\frac{2\kappa p^2}{(\kappa-4)^2}
\end{equation}
and
\begin{equation}\label{xnapp}
\alpha_p=\frac{\kappa-8}{2\kappa}-\frac{p}{2}\;,\quad
\beta_p=\frac{\kappa-8}{2\kappa}+\frac{p}{2}\;,\quad
\gamma_p=\frac{(8-\kappa)^2-\kappa^2p^2}{8\kappa}\;,\quad
x_p=1-\frac{\kappa}{8}+\frac{\kappa}{8}\,p^2\;.
\end{equation}
In order to select the correct set of parameters, it is convenient
to map our problem onto the unit disk $\mathbb{D}$, through the
transformation $z' = \frac{z-w}{z-\bar{w}}\,$ for
$z\in\mathbb{H}\,$ and $z'\in\mathbb{D}$. This transformation maps
the point $w$ to the center of the disk, the length $\epsilon$ to
$\epsilon/|w-\bar{w}|$, and it shifts the angle $\theta$ by an
angle of $\pi/2$. Also, the point $0$ is mapped to $w/\bar{w}$ on
the boundary of the disk, and the point $\infty$ to $1$. We are
then describing an SLE curve on the unit disk started at
$w/\bar{w}$ and required to end at 1. Fixing the power of
$\epsilon/|w-\bar{w}|$ to be some number $x_p$ (the ``scaling
dimension''), we are left, after integration over $\theta$ as in
(\ref{defQn}), with a second order ordinary differential equation
in the angle $\alpha=\arg(w/\bar{w})\in [0,2\pi]$. This equation
is the eigenvalue equation for an eigenfunction of the
two-particle Calogero-Sutherland Hamiltonian with eigenvalue
(energy) $2 x_p/\kappa$ and with total momentum $p$ \cite{johnCS}.
For generic $\kappa$, the Calogero-Sutherland Hamiltonian admits
only two types of series expansions $C\alpha^\omega[[\alpha^2]]$
(with $C\neq0$) as $\alpha\to0^+$ for its eigenfunctions: one with
a leading power $\omega=8/\kappa-1$, the other with a leading
power $\omega=0$. It admits the same two types of series
expansions $C'(2\pi-\alpha)^{\omega'}[[(2\pi-\alpha)^2]]$ (with
$C'\neq0$) as $\alpha\to2\pi^-$. Allowing only one type of series
expansion at $0$ and only one at $2\pi$ (the possibilities give
the Calogero-Sutherland system in the fermionic sector
$\omega=\omega'=8/\kappa-1$, bosonic sector $\omega=\omega'=0$ or
mixed sector, $\omega\neq\omega'$), the Calogero-Sutherland
Hamiltonian has a discrete set of eigenfunctions, with eigenvalues
bounded from below (since it is a self-adjoint operator on the
space of functions with these asymptotic conditions). The lowest
eigenvalue is obtained for the eigenfunction (the ground state)
with the least number of nodes (zeros of the eigenfunction). If
the leading powers $\omega$ and $\omega'$ are chosen equal to each
other, then the ground state (in the sector with total momentum
$n$) is described by the solutions (\ref{Qnsolapp}) with
(\ref{xndiscarded}) (for $\omega=0$) or (\ref{xnapp}) (for
$\omega=8/\kappa-1$), which, in the coordinates of the disk, take
the form
\begin{equation}\label{Qndisk}
Q_p(|w-\bar{w}|,\alpha, \epsilon)\,=\,\tilde{c}_p
\,\left(\frac{\epsilon}{|w-\bar{w}|}\right)^{\,x_p}\,e^{i\,\frac{\alpha_p-\beta_p}{2}\,\alpha}
\,\left(\sin\frac{\alpha}{2}\right)^{\gamma_p+x_p}\;.
\end{equation}
If we consider the probability that the curve passes in between
the two points in the original formulation on the half plane, then
the curve is required here to pass by the center of the disk.
Hence, the probability vanishes when the starting point of the SLE
curve is brought toward its ending point on the disk, from any
direction; this fixes the power to be $8/\kappa-1$ (for
$\kappa<8$) at both values $\alpha=0,2\pi$ and therefore selects
the solution in the fermionic sector (\ref{xnapp}). Note that
since the probability could be given by an excited state in the
fermionic sector (which corresponds to a higher value in place of
the exponent $x_p$), we {\em do not} have the condition that
$\tilde{c}_p$ is nonzero.

\end{appendix}

\end{document}